# Striped-magnetic-order suppresses giant optical anisotropy and drives structural distortion in iron arsenide superconductors

Hai Wang[*]

*Department of Applied Physics, The Hong Kong Polytechnic University, Hong Kong SAR, China*
*College of Materials Science and Engineering, Tongji University, Shanghai 201804, People's Republic of China*

To examine the role of magnetism in superconductivity of iron-based superconductors, we first present first-principles optical calculations on three representative parent compounds: LaFeAsO, BaFe$_2$As$_2$ and LiFeAs. Both nonmagnetic (NM) and spin-density wave (SDW) with striped antiferromagnetic order are considered. Due to their layer structure, these compounds in NM states show anomalously large optical anisotropy with a conductivity ratio $\sigma_{zz}/\sigma_{xx}$ as large as 313% in LaFeAsO and 530% in LiFeAs. Interestingly the giant optical anisotropy is significantly suppressed by SDW. On the other hand, the variable-cell-shape optimizations confirm that SDW does drive an orthorhombic-distortion. The drive force is dependent on the local magnetic moment on Fe sublattice, which is tuned by its bonding environment. Based on these results, we discussed the delicate role of magnetism in superconductivity.

---

[*] corresponding author, email address: hwang@tongji.edu.cn

## I. INTRODUCTION

The discovery of iron-based superconductors (SCs) makes understanding the role of magnetism in superconducting mechanism more important [1]. Generally, magnetism is considered to impede the emergence of superconductivity. For example, the suppression of long-range spin-density-wave (SDW) magnetic order in 1111-type LaFeAsO through chemical-doping [1] [2] or applied-pressure [3] is an effective approach to improving $T_C$ and seems to be the precondition of superconductivity. On the other hand, it is report that the coexistence/competition of magnetism and superconductivity in 122-type $BaFe_2As_2$ [4] [5]. In addition, the SDW is also suggested as the ground-state structure of 111-type NaFeAs systems both experimentally [6, 7] and theoretically [8], as well as predicted in LiFeAs [9] [10]. More and more evidence confirm that superconductivity connects intimately to the SDW magnetic order.

In general, Fe-based superconductors possess a layered tetragonal structure with FeAs carrier conduction layer alternating to the Re-O (or Ba/Li) charge reservoir layer [1] [11]. In fact, the layered structure results in the anisotropy of physical properties with a two-dimensional character [12], such as electronic resistivity [13], thermodynamic and transport properties [14] in $BaFe_2As_2$ systems. Its resistivity anisotropy ($\rho_c/\rho_{ab}$) is found to be as large as 150 [15], more larger than ~50 for $CaFe_2As_2$ [16] and 6.3 for $BaNi_2P_2$ [17]. Similar report for $EuFe_2As_2$ is ~8, which slightly depends on temperature [18]. The superconductivity anisotropy is 3.3 at about Tc for LiFeAs [19].

The intrinsic anisotropy should be not compatible with superconductivity as examined the phase diagram of various iron-based SCs [11]. For example, pure parent compounds 1111-type LaFeAsO and 122-type $BaFe_2As_2$ are not SCs. Both are found to be poor metals at room temperature and undergo a temperature-induced structural transition from tetragonal symmetry to orthorhombic one, followed [2] or accompanied [20] by a magnetic transition from a non-magnetic state to the collinear stripe-type SDW magnetic order. Although similar issue has been given on 111-type NaFeAs systems [6] [7] [8], however, there is no corresponding experimental report for LiFeAs. Different to other SCs, stoichiometric LiFeAs exhibits bulk superconductivity at ambient pressures without chemical doping [21].

While the structural transition is theoretically proposed to be driven by SDW magnetic order [22] [23] [24] [25], the concrete theoretical evidence, however, is still unavailable so far. On the other hand, when temperature lowers than $T_N$, SDW has a significant influence on the optical conductivity in these iron-base SCs. For example, SDW opens an energy gap in LaFeAsO [26], BaFe$_2$As$_2$ [27]. The natural questions are rise up whether SDW control the number of charge carries and its mobility, how SDW affect the anisotropy of electronic, structural and optical properties, and more important one on the role of SDW in superconductivity is still unclear.

In this letter, we present first-principles calculations on the optical properties and structural optimization of three representative parent compounds of iron-base superconductors: LaFeAsO, BaFe$_2$As$_2$ and LiFeAs. It is found that SDW substantially suppress the giant optical anisotropy as well as drive the orthorhombic-distortion, explaining the coexistence of SDW and superconductivity. We focus on the role of magnetic order in the optical properties and structural distortion. These results may improve our understanding on superconducting mechanism.

**II. Method**

All calculations are performed using density functional theory within generalized gradient approximation [28]. The variable-cell-shape structural optimizations on SDW state with $\sqrt{2}a\times\sqrt{2}a\times 1c$ supercell were performed using Quantum-Espresso code [29]. Optical properties were calculated using the OPTIC [30] package of WIEN2k [31]. More details can be found in our work on single-phase multiferroic BiFeO$_3$ [32] and Pb-free ferroelectric BiInO$_3$ [33] and ZnSnO$_3$ [34]. In addition, the intraband contribution of free charge carriers is also taken into consideration. The NM states are calculated using the experimental structure: LaFeAsO (P4/nmm) at 4 K [1], BaFe$_2$As$_2$ (I4/mmm) [20] and LiFeAs (P4/nmm) [21]. To consider the optical properties in SDW state, $\sqrt{2}a\times\sqrt{2}a\times 2c$ supercell is used, hence, the space group is Cccm (66) for BaFe$_2$As$_2$, while Ibam (72) for LaFeAsO and LiFeAs. Here, *a* and *c* are the corresponding experimental lattice constants in NM states. Similar SDW calculations also report for LaFeAsO [10] [35], BaFe$_2$As$_2$ [36] and LiFeAs (AFM-2a state in Ref. [10]). Note that, while LDA calculations give nearly same results and conclusions, using the $\sqrt{2}a\times\sqrt{2}a\times 1c$ supercell [35] does give a different results, but the conclusion that the magnetic order shows a significant affect on the optical properties is also true.

**III. Results and discussion**

For three compounds we considered here, the SDW state is found to be more energetically favorable compared to NM state (Table I), consistent with the ground-state structure report previously. Here, we first discuss the optical properties of LaFeAsO. Both for NM and SDW states, Fig. 1 shows that the low-frequency response is dominated by a narrow Drude component indicating the presence of itinerant charge carriers. Compared to NM, the substantial drop of Drude component suggests that SDW significantly reduces the number of charge carries. This is also consistent with swift reduction of optical conductivity (Fig. 1b). The reflectivity nearly drops linearly with frequency at low- region (< 0.5 eV), then partially restore due to the interband transitions. This agrees well with previous experimental reports [37]. While LiFeAs shows a similar character like LaFeAsO, $BaFe_2As_2$ does not have substantial Drude component at low- region both for NM and SDW states.

Interestingly, anomalously large optical anisotropy is evident for three compounds in NM state (Figs. 1c-d). The conductivity ratio $\sigma_{zz}/\sigma_{xx}$ has a maximum as large as 313% in the range 4-6 eV for LaFeAsO. This is consistent with the suggestion in terms of the DMFT calculation on LaFeAsO [38]. The corresponding value is 231% for $BaFe_2As_2$ and 530% for LiFeAs. For $BaFe_2As_2$, our theoretical value (3.0) at low frequency limit agrees well with experimental report (2.8) [39]. Similar giant optical anisotropy (GOA) has report in $SrFeO_2$ (~220% at 4.9 eV) [40]. The anisotropy is considered to result from the layered-structure and two-dimensional Fermi surfaces along the c-axis [40-43].

More interestingly, GOA is significantly suppressed by SDW (Figs. 1c-d) through, for example in LaFeAsO, increasing $\sigma_{zz}$ in the low-energy range of [0.5, 4] eV as well as decreasing that in high-energy rang [4, 6] eV. Particularly, SDW nearly removes the two sharp peaks in the range of [4, 6] eV of LaFeAsO, while its influence is small on photon energy lager than 6 eV. For $BaFe_2As_2$, SDW also removes the peaks in [3, 6] eV while the optical properties with higher energy remain unchanged. More particularly, SDW nearly takes all the peaks of NM LiFeAs. Whatever, it is confirmed that SDW makes these parent SCs tend to be more isotropic, which is helpful for superconductivity because the two-dimensional superconductivity can be destroyed by a relatively small current [44]. Considered the significant influence of SDW on the optical properties, we suggest that the promising potential application in magnetic-optical devices with excellent tunability maybe found in these tetragonal compound families with a wide range of pnictogens, fluorides and chalcogenides as well as silicide and germanide hydrides.

Experimentally, nearly all optical measurements on iron-based SCs are limited in ab-plane (for c-axis is very rare) and a narrow extent energy range (<4 eV=32400 cm$^{-1}$), maybe because the temperature influence on optical properties is only found in the range of < ~1 eV (8100 cm$^{-1}$). For example, it is reported from 6000 cm$^{-1}$ [37], 3 eV [45, 46], up to 6 eV [26] for LaFeAsO, and 24000 cm$^{-1}$ for NaFeAs [47]. For BaFe$_2$As$_2$, it is from 700 cm$^{-1}$ [48], 1200 [49], 10000 [50], 20000 [39], 24000 [47] to 25000 cm$^{-1}$ [51]. In fact, however, our results indicate that SDW significantly affects the optical properties along zz direction rather than that along xx direction (Fig. 1). Thus, it is highly expected *c-axis* optical measurements on high-quality single-crystals, specifically for higher photon energy at last as large as 6 eV.

For LaFeAsO, compared to NM state, SDW significantly suppresses the reflectivity at low frequencies (< 0.5 eV) (<1200 cm$^{-1}$, see Fig. 3 of Ref. [37]) for both R$_{xx}$ and R$_{zz}$ while only the later in the range of [3.5, 6.4] eV. This results in the reduction of metallicity, *i.e.*, electronic mobility. Our results also reproduce the experimental phenomenon that the optical conductivity are first suppressed and then enhanced in the range of [0.55, 3.1] eV ([700, 2100] cm$^{-1}$ [37]), especially for the zz component [37]. Nearly all Drude component of conductivity is removed, providing the evidence for the formation of SDW gap. Similar spectral features are also report in 122-type AFe$_2$As$_2$ (A=Ba, Sr) [50] and 111-type Na$_{1-x}$FeAs [47]. SDW also results in a large shift of spectral weight from [3.5, 6.4] to [0.55, 3.1] eV in LaFeAsO. This character is also evident for both BaFe$_2$As$_2$ and LiFeAs, indicating generic properties of undoped FeAs-based systems.

The plasma frequencies are calculated (using the method in Ref. [30]) and given in Table I. The results of LaFeAsO agree well with previous report [52]. The calculated value (0.10) of $\left(\omega_{p,xx}^{SDW}/\omega_{p,xx}^{NM}\right)^2$ in BaFe$_2$As$_2$ is also consistent with the experimental data [50]. Compared to NM state, SDW significantly decreases $\omega_{p,xx}$ while increases (or slightly decrease) $\omega_{p,zz}$, resulting in substantial reduction of optical anisotropy, or nearly removing the anisotropy in LiFeAs. As carrier scattering rate (1/$\tau$) possesses same trend [50] with $\omega_p^2$, the reduction of plasma frequencies means the reduction of mobility of free electrons and increase of electronic correlation [53]. On the other hand, the optical spectroscopic combined with band theory through the ratio K$_{exp}$/K$_{band}$ has been used to examine the electronic correlation in 122-type [54] and 1111-type [55] SC, where the electronic correlation is found to be moderate. Because K$_{exp}$/K$_{band}$ is proportional to $\left(\omega_p^{Exp.}/\omega_p^{Cal.}\right)^2$, the electronic correlation evolution from NM to SDW can be examined through calculations on plasma frequency. The fact, that the plasma frequencies of SDW state ($\omega_p^{SDW}$) substantially less than those of NM state ($\omega_p^{NM}$)), suggests SDW significantly increases the electronic correlation, which should impede superconductivity if it is strong as large as to drive an orthorhombic-distortion (later shown). This seems to be in contrary with the fact that SDW can reduce the optical anisotropy, which is helpful for SC.

In addition, it is evident that SDW dramatically reduces $1/\tau$ [50]. To examine the influence of $1/\tau$ on optical properties, using the NM LaFeAsO as a sample, we present the reflectivity and absorption spectroscopy (Fig. 2) as a function of $\sigma^D$, which is nearly in inverse proportion to $1/\tau$. Interestingly $\sigma^D$ affects only the Drude component of optical properties, while the others remain unchanged especially when photon energy >1 eV. As $\sigma^D$ increases, i.e., $1/\tau$ decrease, the reflectivity in xx(zz) direction first decreases in the range of [0,0.4] ([0,0.1]) eV, then increases in the range of [0.4,1] ([0.1,0.5]) eV. Consequently, $\sigma^D$ solely reproduces the character of optical conductivity reported experimentally. Hence, Fig. 2 provides concrete evidence that substantial reduction of carrier scattering rate results in the formation of the SDW gap.

Now, we turn to the influence of SDW magnetic order on the structural transition through full variable-cell-shape optimizations on three SCs. The lattice constants (*a*, *b*) are found to split from the same to different (Fig. 3b), here, lattice constant *a* is compressed and *b* is elongated while the angle between them are unchanged, resulting in an orthorhombic structural distortion. Meanwhile, the total energy of system decreases (Fig. 3a), suggesting the orthorhombic–distortion from tetragonal parent phase is energetically favorable (~1.7 mRy). The SDW driven orthorhombic–distortion is also evident for both $BaFe_2As_2$ and LiFeAs. To examine whether SDW drives the distortion and whether its drive force connects to the local magnetic moment of Fe sublattice ($MM_{Fe}$), we performed fixed spin moment (FSM) calculations with constrained $MM_{Fe}$. Results show that the less $MM_{Fe}$ the less lattice constants (*a*, *b*) (Fig. 4d), providing unambiguous evidence for $MM_{Fe}$ controls the distortion magnitude, i.e., SDW drives the orthorhombic-distortion. When $MM_{Fe}$ takes as 38% of theoretical value, volume collapse (~3%) is observable (Fig. 4c). In addition, the La-z and As-z atomic fractional coordinations (Fig. 4e, f) are also increase 2.1% and 2.8%, respectively. This suggests that $MM_{Fe}$ also controls the height of anion (here, it is As) as well as both the Fe-As bond-length and the inter-layer distance between La-O and Fe-As layers. Note that the pnictogen height ($h_{Pn}$) is considered as a key factor that is correlated to the Tc of Fe-based SCs [56]. Here, our results show that $h_{Pn}$ is controlled by $MM_{Fe}$. Therefore, the values of local

MM$_{Fe}$, i.e., the local-moment interaction is a dominant factor over the Fermi-surface nesting in determining the stability of the magnetic phase and that the partial gap is an induced feature by a specific magnetic order [57]. For LiFeAs, the $\triangle E$ is about 35 mRy, about two times larger than that in LaFeAsO. However, SDW drives structural transition has not been found in LiFeAs systems, indicating more large drive force is need to drive an orthorhombic-distortion than that in LaFeAsO systems. Further systemic studies on how to control SDW to improve T$_C$ in 1111/122/111-type iron-base SC are high expected.

## IV. Conclusions

Based on first-principles calculations, we present the influence of SDW magnetic order on the optical properties and structural distortion in three iron-based SCs. The results show that these compounds in nonmagnetic state possess giant optical anisotropy, which is significantly suppressed by SDW. To confirm this, it is highly expected that further optical experiments on the high-quality single-crystal for photon energy at last as large as 6 eV. On the other hand, SDW drives an orthorhombic-distortion and introduces in-plane structural anisotropy. The driven force is closely connected to MM$_{Fe}$, which is tunable through chemical-doping or applied-pressure. Hence, if MM$_{Fe}$ is large enough to drive a structural transition, SDW impedes superconductivity. If it not, SDW is helpful for superconductivity owing to suppression of giant anisotropy. Therefore, further study on how to control the value of MM$_{Fe}$ to improve T$_C$ is also highly expected.

TABLE I. Calculated plasma frequencies (eV) and the total energy difference $\Delta E$ (eV/unit cell) compared to NM.

|  | LaFeAsO | | BaFe$_2$As$_2$ | | LiFeAs | |
|---|---|---|---|---|---|---|
|  | NM | SDW | NM | SDW | NM | SDW |
| $\Omega_{p,xx}$ | 2.22 (2.3[a]) | 0.69 (0.67[b]) | 2.63 | 0.80 | 2.33 | 0.96[c] |
| $\Omega_{p,zz}$ | 0.45 (0.32[a]) | 0.85 | 0.49 | 0.44 | 1.86 | 1.15[c] |
| $\Delta E$ | 0 | -0.91 | 0 | -1.30 | 0 | -1.42 |

[a] Reference [52]. [b] Reference [26]. [c] For LiFeAs: $\Omega_{p,xx}$ 0.9183 (up) 1.01 (dn), and $\Omega_{p,zz}$ 1.1105 (up) 1.19 (dn). The values for SDW are their average.

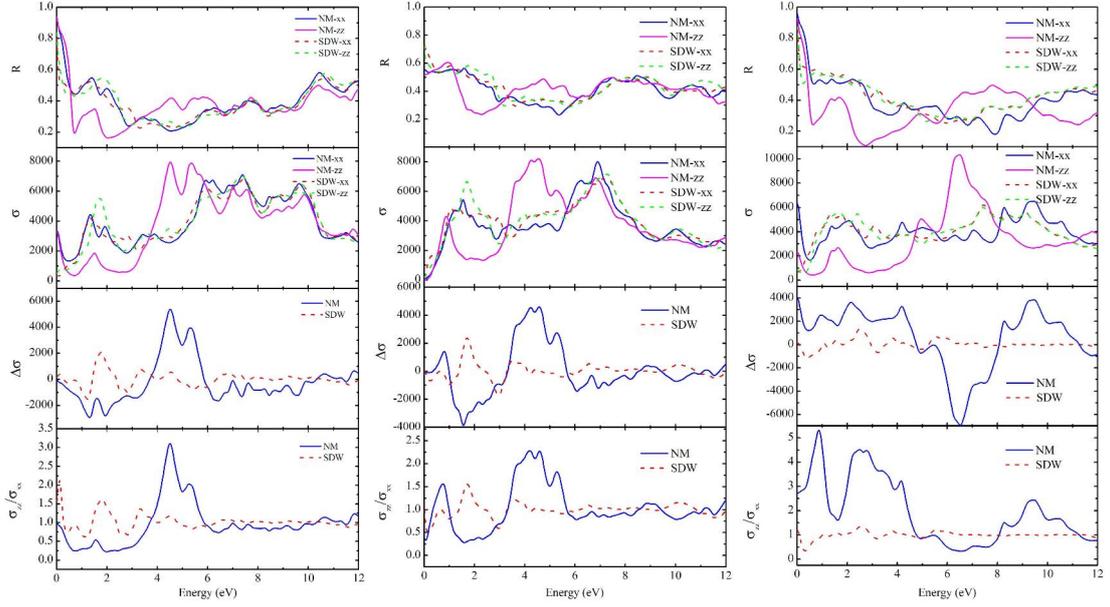

FIG. 1 (Color online). Calculated optical properties of LaFeAsO (left), BaFe$_2$As$_2$ (center) and LiFeAs (right) for both NM and SDW states. We show the reflectivity (a), optical conductivity (1/Ω cm$^{-1}$) (b), difference conductivity Δσ = σ$_{zz}$- σ$_{xx}$ (c), and the conductivity ratio σ$_{zz}$/σ$_{xx}$ (d). Giant optical anisotropy in the NM state is evident in panel (d), where the anisotropy is also significantly suppressed by SDW.

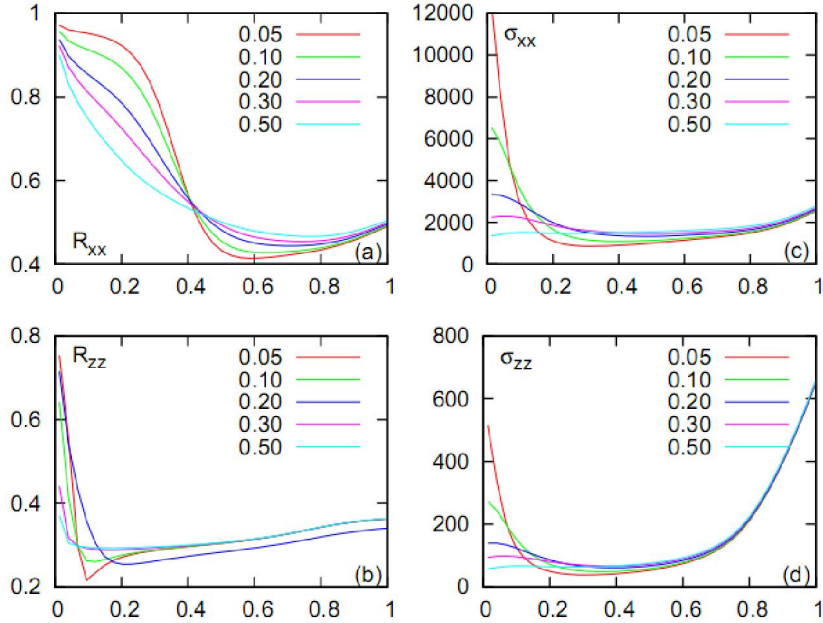

FIG. 2 (Color online). Calculated reflectivity (left) and absorption (right) of NM LaFeAsO along both xx (a,c) and zz (b,d) directions as a function of ω$^D$, which controls the Drude component of optical properties and nearly no influence on these large than 1 eV.

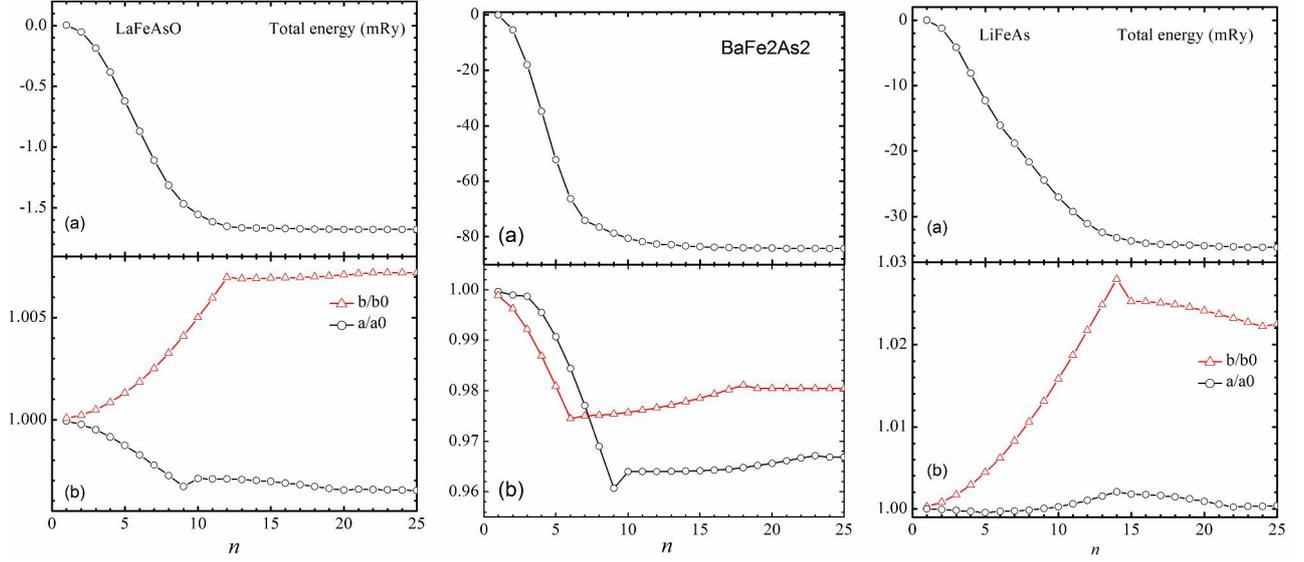

FIG. 3 (Color online). SDW driven orthorhombic structural distortion in LaFeAsO (left), BaFe$_2$As$_2$ (center) and LiFeAs (right). The total energy (a) and normalization lattice constant ($a/a_0$, $b/b_0$) (b) as a function of optimization step ($n$).

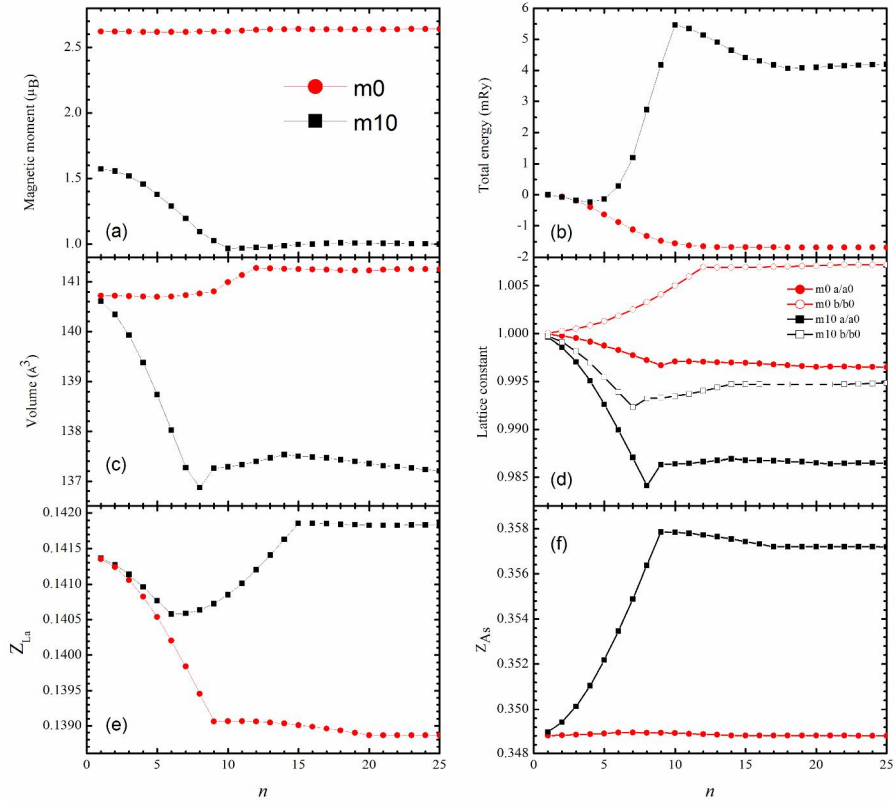

FIG. 4 (Color online). FSM calculation shows the MM$_{Fe}$ (a), total energy (b), volume (c), normalization lattice constants (d) and the atomic fractional coordinations of La-z (e) and As-z (f) in LaFeAsO. Here, m0 and m10 stand for zero and 10% constrain of magnetization/charge, respectively.